\documentclass{article}
\usepackage{geometry}
\usepackage[utf8]{inputenc}
\usepackage{graphicx}
\usepackage{amsmath}
\usepackage{authblk}
\usepackage[english]{babel}
\usepackage{xcolor}

\newcommand\red[1]{{\color{black}#1}}

\title{Neuronal growth on high-aspect-ratio diamond nanopillar arrays for biosensing applications}

\author[1]{Elena Losero}
\author[2]{Somanath Jagannath}
\author[2]{Maurizio Pezzoli}
\author[3]{Valentin Goblot}
\author[3]{Hossein Babashah}
\author[2]{Hilal A. Lashuel}
\author[3]{Christophe Galland}
\author[4]{Niels Quack}

\affil[1]{Division of Quantum Metrology and Nanotechnologies, Istituto Nazionale di Ricerca Metrologica (INRiM), Strada delle Cacce 91, 10135 Torino, Italy}
\affil[2]{School of Life Sciences, EPFL, Rte Cantonale, 1015, Lausanne, Switzerland}
\affil[3]{School of Basic Sciences, Institute of Physics, EPFL, Rte Cantonale, 1015, Lausanne, Switzerland}
\affil[4]{School of Aerospace, Mechanical and Mechatronic Engineering, The University of Sydney, NSW, Australia}
\date{}

\begin{document}

\maketitle

\begin{abstract}
    Monitoring neuronal activity with simultaneously high spatial and temporal resolution in living cell cultures is crucial to advance understanding of the development and functioning of our brain, and to gain further insights in the origin of brain disorders. 
    While it has been demonstrated that the quantum sensing capabilities of nitrogen-vacancy (NV) centers in diamond allow real time detection of action potentials from large neurons in marine invertebrates, quantum monitoring of mammalian neurons (presenting much smaller dimensions and thus producing much lower signal and requiring higher spatial resolution) has hitherto remained elusive. In this context, diamond nanostructuring can offer the opportunity to boost the diamond platform sensitivity to the required level. However, a comprehensive analysis of the impact of a nanostructured diamond surface on the neuronal viability and growth was lacking. Here, we pattern a single crystal diamond surface with large-scale nanopillar arrays and we successfully demonstrate growth of a network of living and functional primary mouse hippocampal neurons on it. Our study on geometrical parameters reveals preferential growth along the nanopillar grid axes with excellent physical contact between cell membrane and nanopillar apex. Our results suggest that neuron growth can be tailored on diamond nanopillars to realize a nanophotonic quantum sensing platform for wide-field and label-free neuronal activity recording with sub-cellular resolution.
\end{abstract}




\begin{figure}[htb]
    \centering
    \includegraphics[width=0.9\textwidth]{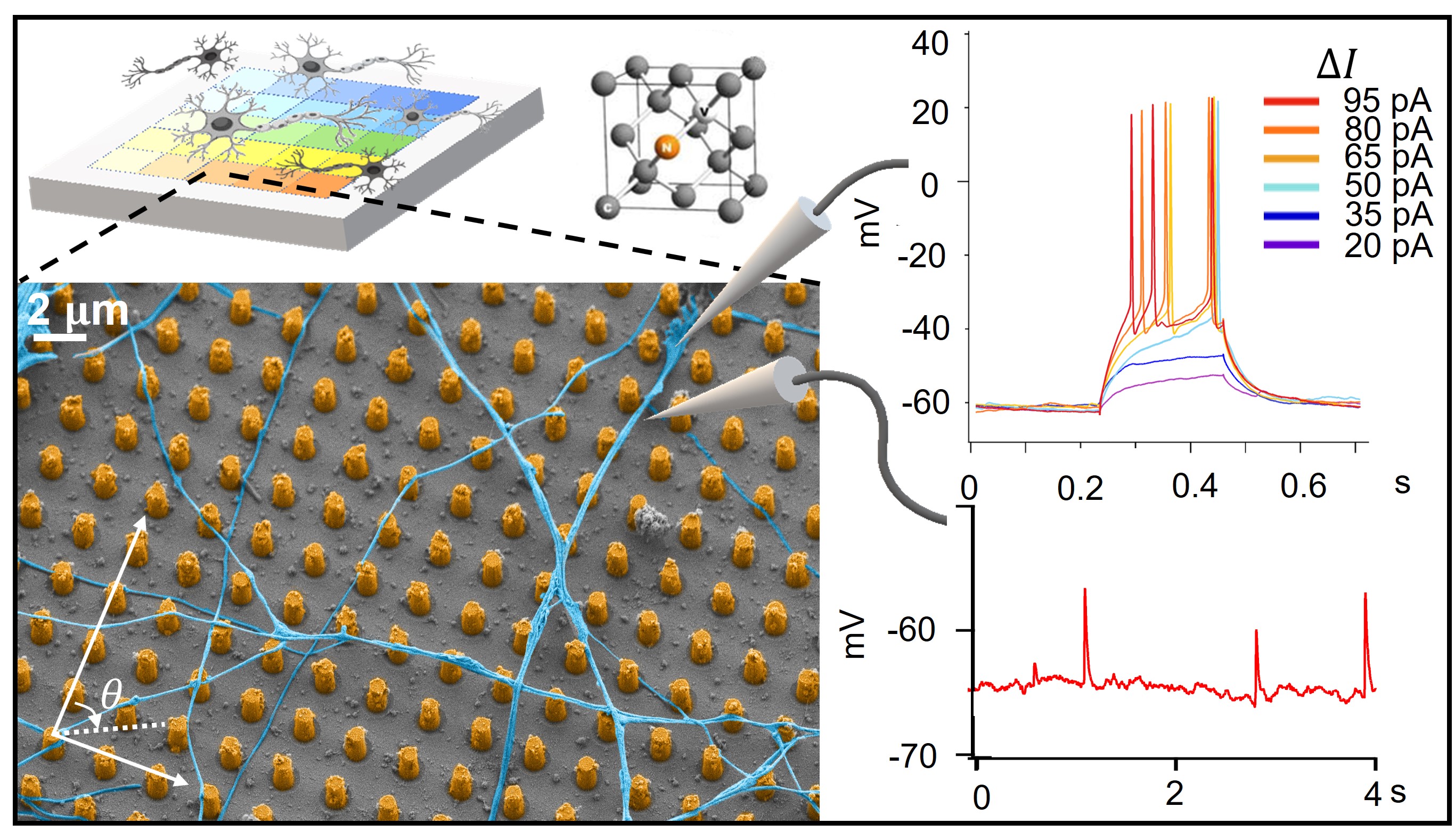}
    \end{figure}

\section*{Introduction}

Despite tremendous progress in neuroscience, a series of fundamental questions about the operation of the brain and the origin of neurodegenerative diseases remain unsolved \cite{adolphs2015unsolved,lee2011protein}. 
One of the most powerful avenues to advance our understanding of the human brain is the analysis of the \red{electro-chemical} signals between individual neurons. An ideal recording instrument would provide minimally invasive access to highly connected networks of neurons with sub-cellular spatial resolution and \red{sub-millisecond temporal resolution}.    
Commonly employed techniques for recording of neuronal activity include patch clamp \cite{neher1976single,hamill1981improved,neher1992patch}, multi-electrode arrays \cite{thomas1972miniature,gross1982recording,spira2013multi}, genetically encoded voltage or calcium indicators \cite{siegel1997genetically,miyawaki1997fluorescent,persechini1997novel,yang2016genetically} or optogenetic tools \cite{fenno2011development,yizhar2011optogenetics,packer2013targeting}. However, these techniques are relatively invasive, allow simultaneous monitoring of only a small number of neurons, and typically exhibit poor spatial and temporal resolution \cite{hong2019novel}. 
Overcoming these limitations would permit deeper understanding of the dynamics of neuronal networks and of information processing in our brain.
Moreover, measuring with high sensitivity the electric and magnetic fields generated by neuronal activity can also be useful for identifying early stages of brain disorders such as Parkinson's and Alzheimer's \cite{sakmann1995single,waxman1998demyelinating} and for better modelling of neuronal biophysics \cite{sporns2014contributions}.

Diamond has been recognized as an ideal candidate material for novel biosensing platforms \cite{wu2016diamond,schirhagl2014nitrogen,zhang2021toward,petrini2020quantum,troise2022vitro} as it is biocompatible \cite{yu2005bright} and features outstanding chemical and physical properties \cite{mildren2013optical}.
For example, diamond has been used as neural interface for electrodes and microelectrode arrays \cite{garrett2016diamond,yang2015diamond}. 
Less explored in the context of neuroscience are the intrinsic defect centers that diamond contains and that can be used for remote sensing of magnetic and electric field at sub-micrometer resolution.
In particular, diamond can host nitrogen-vacancy (NV) centers consisting of a nearest-neighbor pair of a nitrogen atom (substituting for a carbon atom) and a lattice vacancy.

\red{NV centers in diamond feature outstanding sensing capabilities.Optical polarisation and readout of their spin states by optically detected magnetic resonance (ODMR) has been demonstrated to be a powerful tool for sensing magnetic fields, but also for electric fields, temperature and strain \cite{rondin2014magnetometry,jensen2017magnetometry,radtke2019nanoscale,rembold2020introduction}. Moreover, coherence time monitoring and spin relaxometry led to great results in concentration measurements of paramagnetic ions and molecules in
solution" \cite{barton2020nanoscale,perona2020nanodiamond}. Finally, the presence of different charge states of the center (i.e. $NV^+, NV^0, NV^-$) with different photoluminescence spectra can be used to monitor voltage changes in solutions \cite{krecmarova2021label,karaveli2016modulation} and has recently been shown to allow for detection of biologically relevant signal magnitudes \cite{mccloskey2022diamond}.}
Both nanodiamonds and bulk diamond have been used for NV-based sensing. 
Although nanodiamonds are easier to implement in a biological environment and have been successfully employed in biosensing studies \cite{mcguinness2011quantum,nie2021quantum,yukawa2020quantum,hsiao2016fluorescent,chipaux2018nanodiamonds} their NVs are randomly oriented, which prevents the implementation of many sensing schemes that require precisely oriented bias magnetic fields (such as for electric field sensing \cite{dolde2011electric,michl2019robust}) and they generally feature shorter coherence times.
On the contrary, near-surface NV centers in single-crystal bulk diamond can retain excellent quantum coherence after appropriate surface functionalization for biosensing \cite{xie2022biocompatible}.

An analytical model to estimate the magnetic and electric fields experienced by an NV center near a typical mammalian neuron was presented by Hanlon \textit{et al.} \cite{hanlon2020diamond}, showing that for neurons on a flat diamond substrate the required sensitivity is beyond what has been achieved so far using ODMR techniques. However, they argued that this challenge can be overcome with a nanostructured diamond surface. In particular, if the diamond surface is structured as a nanopillar array, on which the neurons are grown, the direct contact between diamond nanopillars and cell membrane avoids the screening of local electric fields by the surrounding ionic medium. \red{In the same work, neurons are grown on the structured surface but their physiological functionalities as well as the relative position between the neurons and the nanostructures at the nanoscale are not explored.} 

In earlier works the authors measured the magnetic fields generated by very large axons from invertebrate species \cite{hall2012high,barry2016optical}  (e.g marine worm and squid, axon diameter up to 1.5~mm \cite{ortega2020neurons}). 
Mammalian neurons present much smaller dimensions (axon diameter typically lower than $1~\mu \mathrm{m}$) and generate much weaker magnetic and electric fields, 
\red{requiring the development of new, more sensitive techniques. In \cite{mccloskey2022diamond}, a diamond voltage microscope was demonstrated that is able to detect an \emph{artificial} signal mimicking the one generated by mammalian neurons. This device is based on the controlled preparation of the NV charge state by surface functionalization. The substrate consists of an ultra pure single-crystal diamond presenting a dense layer of NV centers close to the surface. Nanostructuring with diamond pillar arrays is used to enhance the collection efficiency, and thus the sensitivity, preserving high spatial resolution and wide field of view \cite{momenzadeh2015nanoengineered,hanlon2020diamond,fuchs2018optimized}. Note that, despite the biosensing application in mind, no neurons were grown on the substrate in \cite{mccloskey2022diamond}.}

Despite steady progress in micro- and nanostructuring of single crystal diamond, reliable and scalable fabrication processes for diamond photonics platforms and biosensing applications are challenging and remain in the development stage \cite{mi2020integrated}. Moreover, even if diamond biocompatibility has already been demonstrated, no systematic study of neuronal activity and morphology has been performed on cells grown on nanostructured diamond and investigation of the neuron-diamond interface with nanoscale spatial resolution remains elusive. 

Here, we demonstrate a highly controllable and repeatable nanofabrication process for producing high aspect ratio, large scale nanopillar arrays made of single crystal diamond. 
\red{To the best of our knowledge, these are the largest and most uniform diamond nanopillar arrays demonstrated today. }
Taking advantage of the NVs naturally present in our sample, we verify enhanced photoluminescence from pillar locations while spin resonance spectra remain unaffected by the nanostructuring. 
In addition, we thoroughly characterize the neurons grown on the nanostructured diamond surfaces. 
\red{We demonstrate with electrophysiology measurements that the neurons are alive and functional, with no obvious alteration of their activity compared to reference substrates.}
Finally, we investigate the relative position between the cells and the pillars \red{at the nanometer scale.} 
\red{Using scanning electron microscopy (SEM) we establish how neurons grow depending on varying geometrical parameters of the array. In particular, we show that for a suitable set of parameters neurons can successfully grow suspended on top of the pillars.}
\red{Our findings are of prime relevance for NV sensing schemes such as the ones suggested in \cite{mccloskey2022diamond, hanlon2020diamond}, where it is assumed that the cells are in close proximity with the NV centers at the pillar apex.} Our quantitative experimental studies of neuronal growth on diamond nanopillars \red{will guide precise estimate of electric field and/or voltage expected at the NV center position}. 
\red{We also explore the impact of our diamond nanostructures in guiding the neuronal growth into preferential directions. Previous studies reported on the effect of micro- and nanostructured surfaces on neuronal growth for different substrate materials, topography or cell types \cite{gautam2017engineering,marcus2017interactions,simitzi2017controlling,martinez2009effects,hoffman2010topography,repic2016characterization,dowell2004topographically,specht2004ordered}. However, due to the differences in the nanofabrication processes and resulting surface properties, the findings from other materials are not immediately transferable to diamond, and our dedicated experiments fill this gap.}

\section{Results and discussion}

\subsection{Large-scale high aspect-ratio nanopillar arrays}

\begin{figure}
    \centering
    \includegraphics[width=0.9\textwidth]{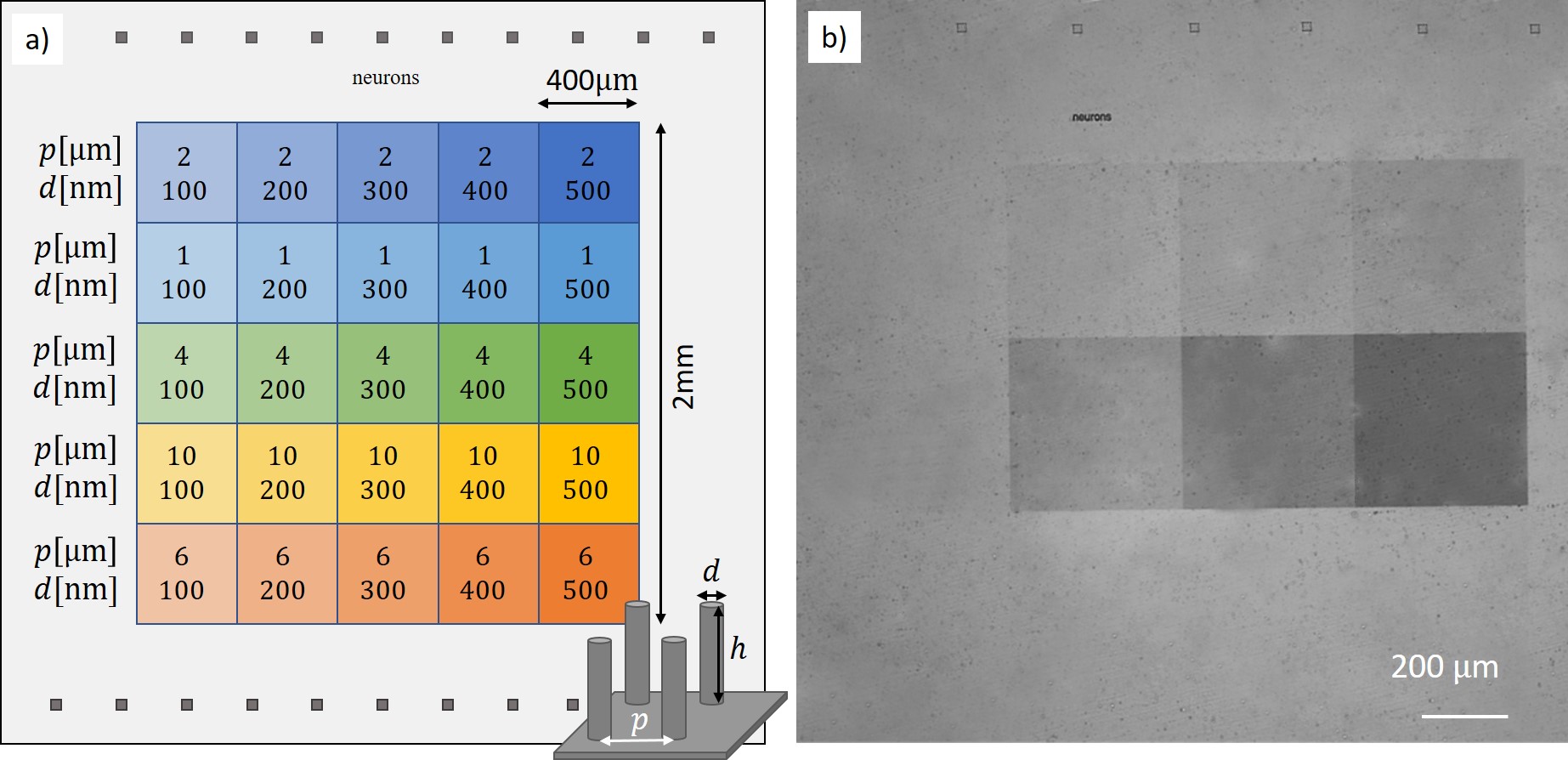}
    \caption{a) Design of 25 nanopillar arrays with different geometrical parameters on the same chip, each array covering an area of $400~\mu\mathrm{m}~\times~400\mu\mathrm{m}$. They all present the same height $h \sim 1\mu$m. b) Optical image of the diamond surface before plating the neurons. The different pillar densities and diameters yield different colors. Pillars with diameters $d\leq 300$~nm and/or pitch $p \geq 2 \mu$m cannot be clearly distinguished.}\label{fig:Design}
\end{figure}

Following our optimized nanofabrication protocol (cf. Methods section) we produce large-scale arrays of nanopillars with diameters as small as 100~nm and uniformly covering areas up to 2~mm~$\times$~2~mm on the surface of a single crystal CVD diamond substrate of 3~mm~$\times$~3~mm~$\times$~0.3~mm in size. We divide the surface in 25 arrays with varying pitch $p$ along rows (1, 2, 4, 6, and 10~$\mu$m), and varying pillar diameter $d$ along columns (100, 200, 300, 400, and 500~nm), respectively. The design of the nanopillar arrays is schematically represented in Fig.~\ref{fig:Design}a and a microscope recording of a fabricated large-scale nanopillar array is presented in Fig.~\ref{fig:Design}b.
Note that these dimensions are not technological limits, yet rather design choices: for applications requiring further reduction of the pillar diameter, the e-beam lithography process can be further optimized, and the array size is limited by the substrate size and can be extended on suitable larger single crystal diamond substrates.

\begin{figure}
    \centering
    \includegraphics[width=1\textwidth]{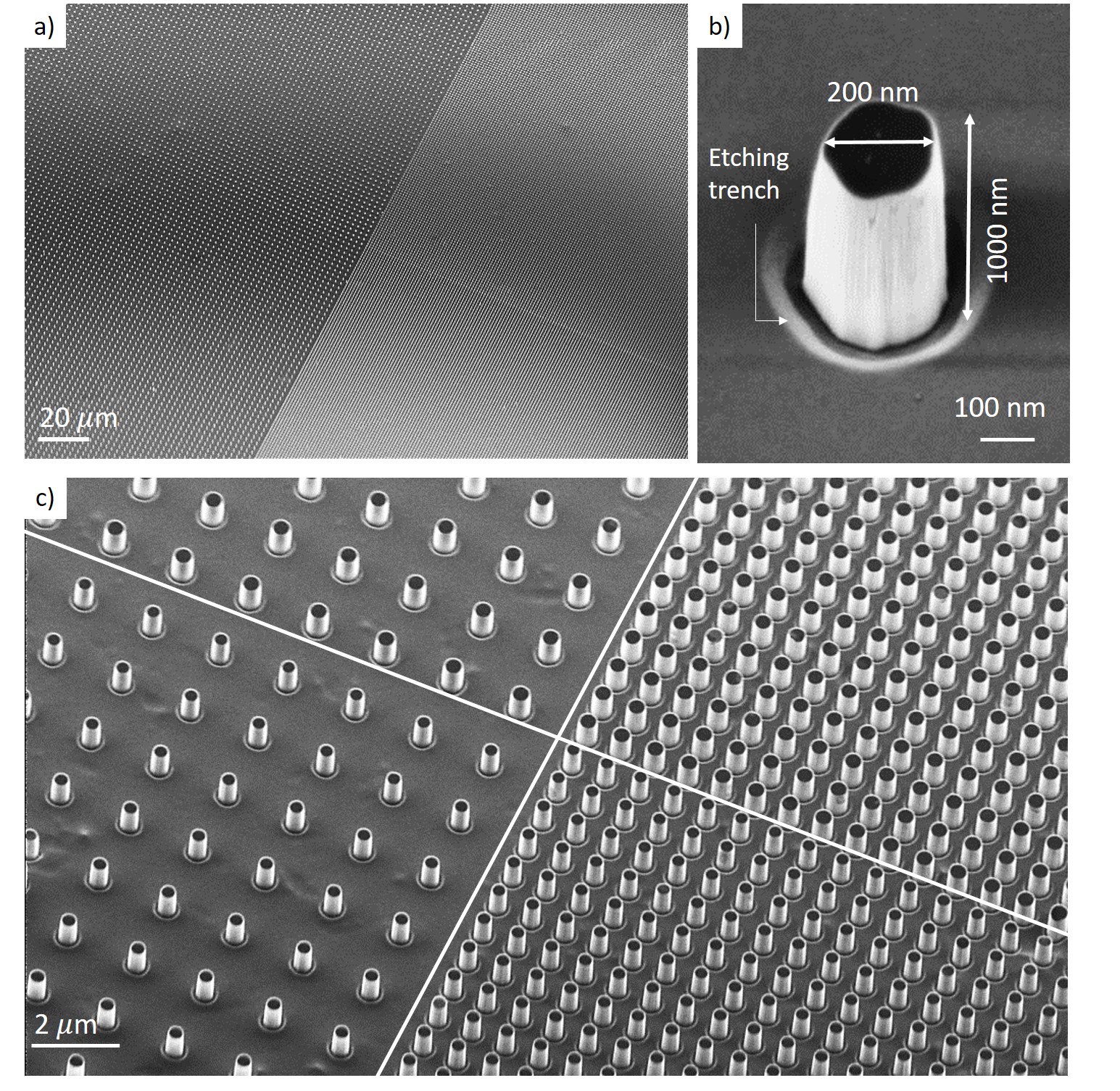}
    \caption{a-c) SEM images of adjacent diamond pillar square arrays. They reveal high homogeneity and smooth and clean diamond surface between the pillars. b) SEM image of an individual diamond pillar, with a diameter of 200~nm and a height of 1~$\mu$m. As a result of the optimized etching process, the top is flat and the sidewalls exhibit high verticallity.}\label{fig:Substrate}
\end{figure}

Our optimized nanofabrication protocol allows to obtain arrays with excellent uniformity as shown in the wide field scanning electron microscope recording of the intersection between four selected arrays in Fig.~\ref{fig:Substrate}a. Owing to the highly directional oxygen plasma etch process employed, the individual pillars present vertical sidewalls and excellent aspect ratios exceeding 1:10, while keeping a flat top, as shown in Fig.~\ref{fig:Substrate}b. A defect-free and smooth bottom surface stretches between the pillars over the entire surface, as shown in Fig.~\ref{fig:Substrate}c. 

\begin{figure}
    \centering
    \includegraphics[width=1\textwidth]{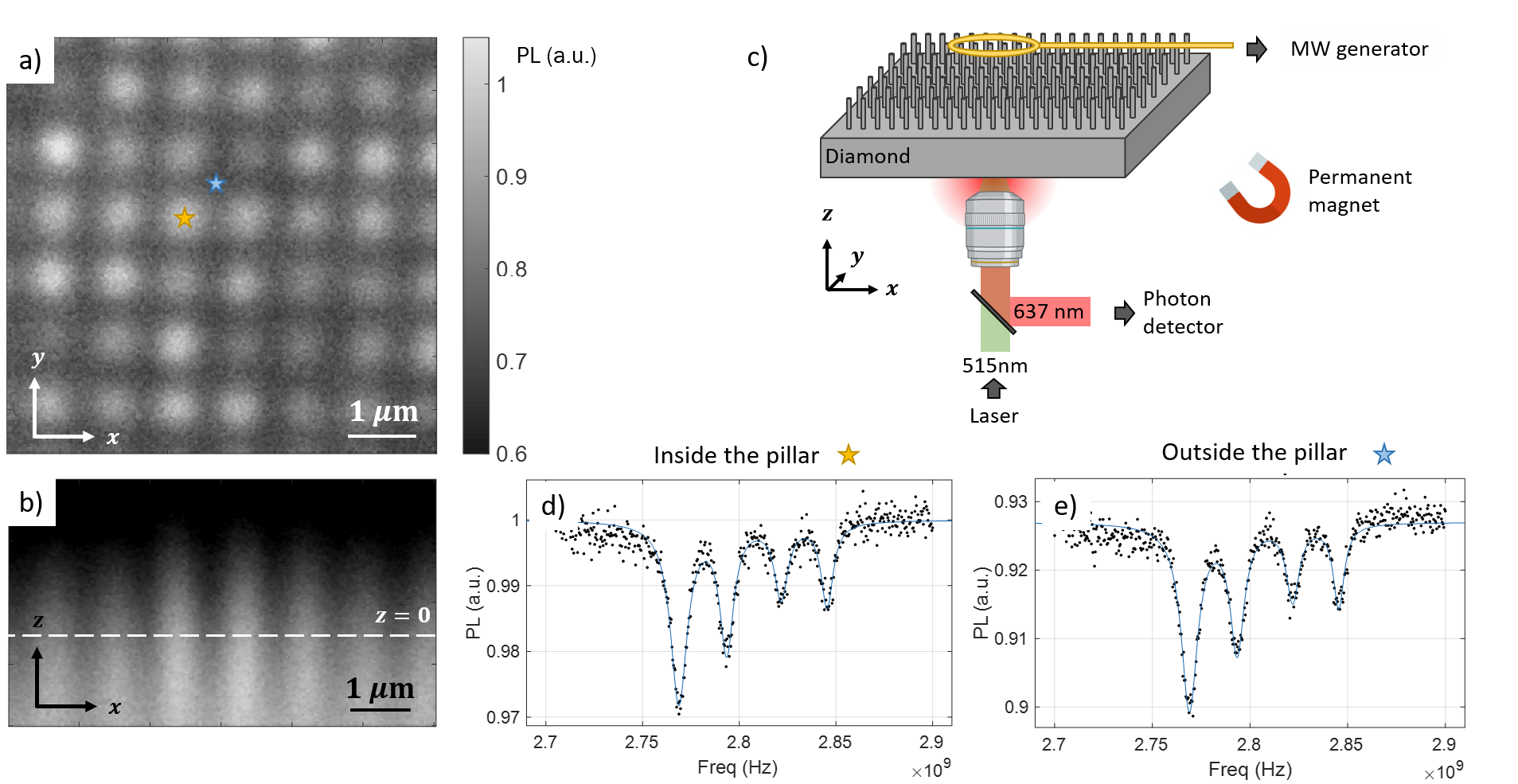}
    \caption{a) Photoluminescence $xy$-map at the diamond surface level ($z=0$). The pillars ($d=500$~nm and $p=1\mu$m) can be clearly distinguished as brighter spots. b) Photoluminescence $xz$-map, demonstrating the pillar waveguiding effect. The dashed line represents the estimated surface level. c) Schematic of the experimental configuration used for both PL and ODMR measurements. Figure not on scale. d) Continuous-wave ODMR spectrum acquired from one pillar, in presence of a magnetic field (few mT). For better visibility only the lower half of the spectrum is reported. The other four resonances are symmetric with respect to the zero-field resonance at 2.87~GHz. e) Continuous-wave ODMR spectrum acquired in between pillars, in the same experimental conditions as for the spectrum in d). \red{All the plots are normalized to the largest PL signal level measured on this area.} }\label{fig:PL}
\end{figure}

\red{In an NV-sensing platform, efficient collection of the photoluminescence (PL) signal from the color centers is essential}.
In order to verify that nanopillars can funnel PL to the back of the substrate \cite{babinec2010diamond,hedrich2020parabolic,momenzadeh2015nanoengineered,hanlon2020diamond}, we mapped the PL intensity collected under our home-built confocal microscope by scanning the sample in the $xy-$ and $xz-$ planes respectively,  Fig.~\ref{fig:PL}a-c. Exciting and collecting PL from the back of the substrate represents the most convenient configuration when dealing with biological samples, also leaving the top surface available for complementary probes such as microelectrodes. 
Even though the substrate used for this experiment contains uniformly distributed NV centers (concentration estimated as 1.4 ppb), we observe that the collected PL intensity is higher when focusing on a pillar ($+15\%$ compared to flat surface). In particular, Fig.~\ref{fig:PL}b shows how the light is guided along the nanostructures. 
Larger enhancements have been shown possible using optimal pillar geometries \cite{hedrich2020parabolic} or pyramids \cite{batzer2020single}. 

In Fig.~\ref{fig:PL}d,e we report the continuous-wave ODMR spectra from a region inside and outside a pillar, respectively, in the presence of a magnetic field ($\sim 3-4$ mT) to split the degeneracy. Except for the higher PL intensity, no significant difference is visible in terms of linewidth and contrast, demonstrating that we can successfully extract ODMR signal from NV centers inside the nanostructrues. \red{This represents an essential element toward the development of a sensing protocol based on optical readout of the spin states.} 
\red{More detailed characterization of the effects of nanostructuring on NV center properties such as on their coherence times can be found in other works, for example Ref.~\cite{volkova2022optical,mccloskey2020enhanced}.}

\subsection{Neuronal morphology}

\begin{figure}
    \centering
    \includegraphics[width=1\textwidth]{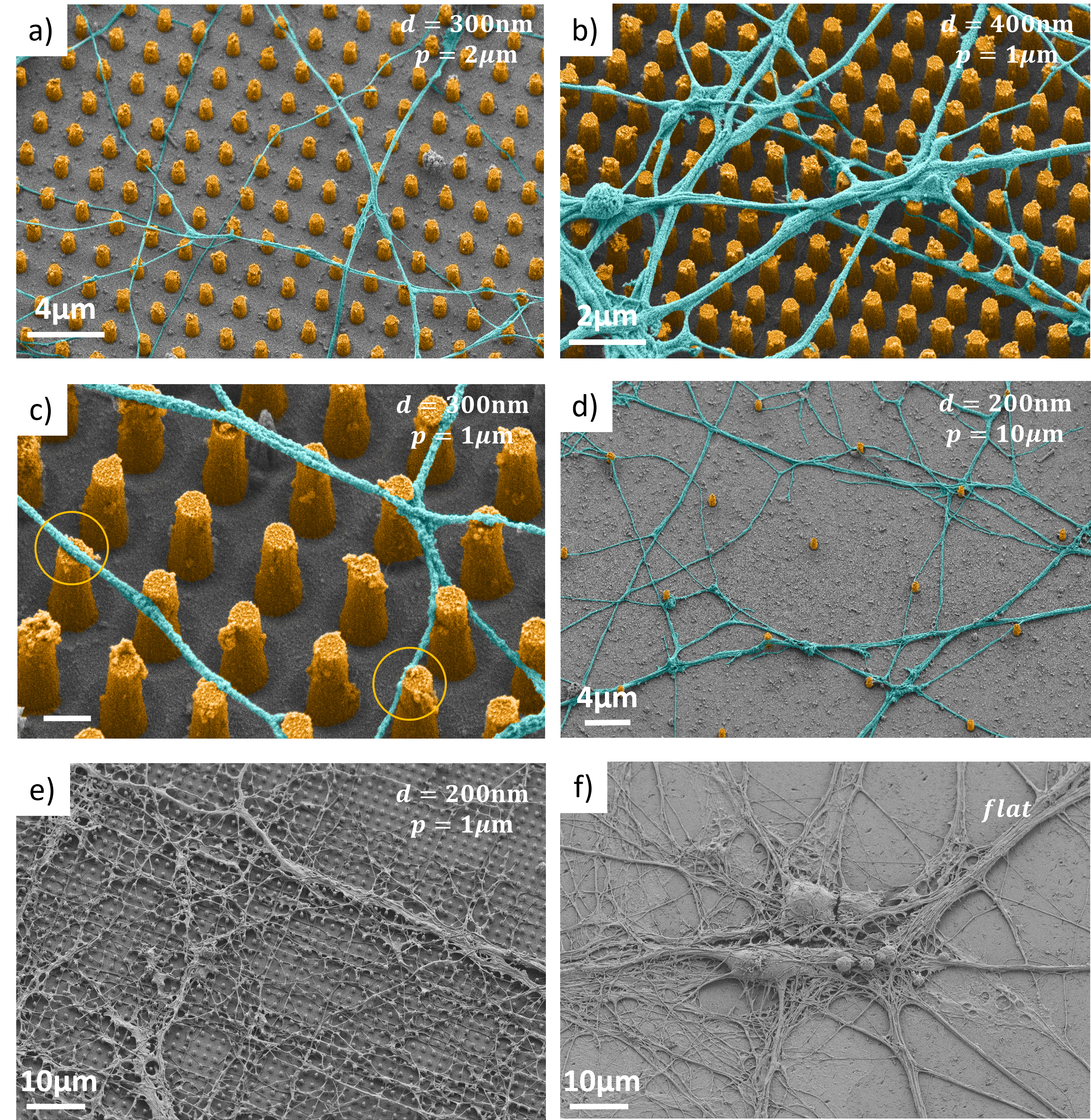}
    \caption{Axon-pillar relative positions. SEM images \red{of primary hippocampal neurons} plated on different diamond substrates, at magnification up to $\times$50k. The samples are tilted by 20$^\circ$ with respect to horizontal to better appreciate the three-dimensional morphology. In panels~a-e the substrate is nanostructured with pillar arrays (height $\sim 1\mu$m, pitch $p$ and diameter $d$ reported in each figure). In panel~f the substrate is flat. Images a-d have been artificially colored for better visibility. }\label{fig:SEMneuronsposition} 
\end{figure}

Figures~\ref{fig:SEMneuronsposition}a-e present SEM recordings of neurons grown on diamond nanopillar arrays with different geometrical parameters, at different magnification factors (up to $\times$50k).
From Fig.~\ref{fig:SEMneuronsposition}a-c, we can appreciate that almost all neurites are suspended over the substrate and run on top of the pillars for pitches of $1-2~\mu\mathrm{m}$. Fig.~\ref{fig:SEMneuronsposition}c reveals two common configurations for the relative position between neurites and pillars: either precisely on the top (circle on the left) or on the side, close to the top (circle on the right).
We observe that a majority of \red{neurites} are suspended as long as $p \leq 4\mu\mathrm{m}$, irrespective of the pillar diameter in the tested range.
As the distance between pillars increases (e.g. $p=10~\mu\mathrm{m}$, Fig.~\ref{fig:SEMneuronsposition}d),
the number of suspended \red{neurites} decreases and an increasing number of neurons grows on the diamond floor rather than on top of nanopillars, without being in contact with the pillars. 

The effect of pillar diameter on neuron growth is less evident, at least for the limited range experimentally explored. 
However, we note that the \red{neurites} tend to grow preferentially on the nanopillar top surface rather than on the nanopillar side when the pillar diameter is large with respect to the distance between the pillars (e.g. for $p=1~\mu\mathrm{m}$ and $d \geq 300\mathrm{nm}$). 

The SEM recordings further reveal  a  close  contact  of the \red{neurites} with  the  diamond pillars,  without  evidences  of membrane perforation. 
In Fig.~\ref{fig:SEMneuronsposition}f, an image of a neuron on the flat surface is reported and can be compared with Fig.~\ref{fig:SEMneuronsposition}e acquired with the same magnification factor but on a pillar array. 
The structured substrates do not significantly alter the neuron morphology, which is in agreement with the well preserved functional properties, as demonstrated by electrophysiology (see next section). 

\begin{figure}
    \centering
    \includegraphics[width=1.1\textwidth]{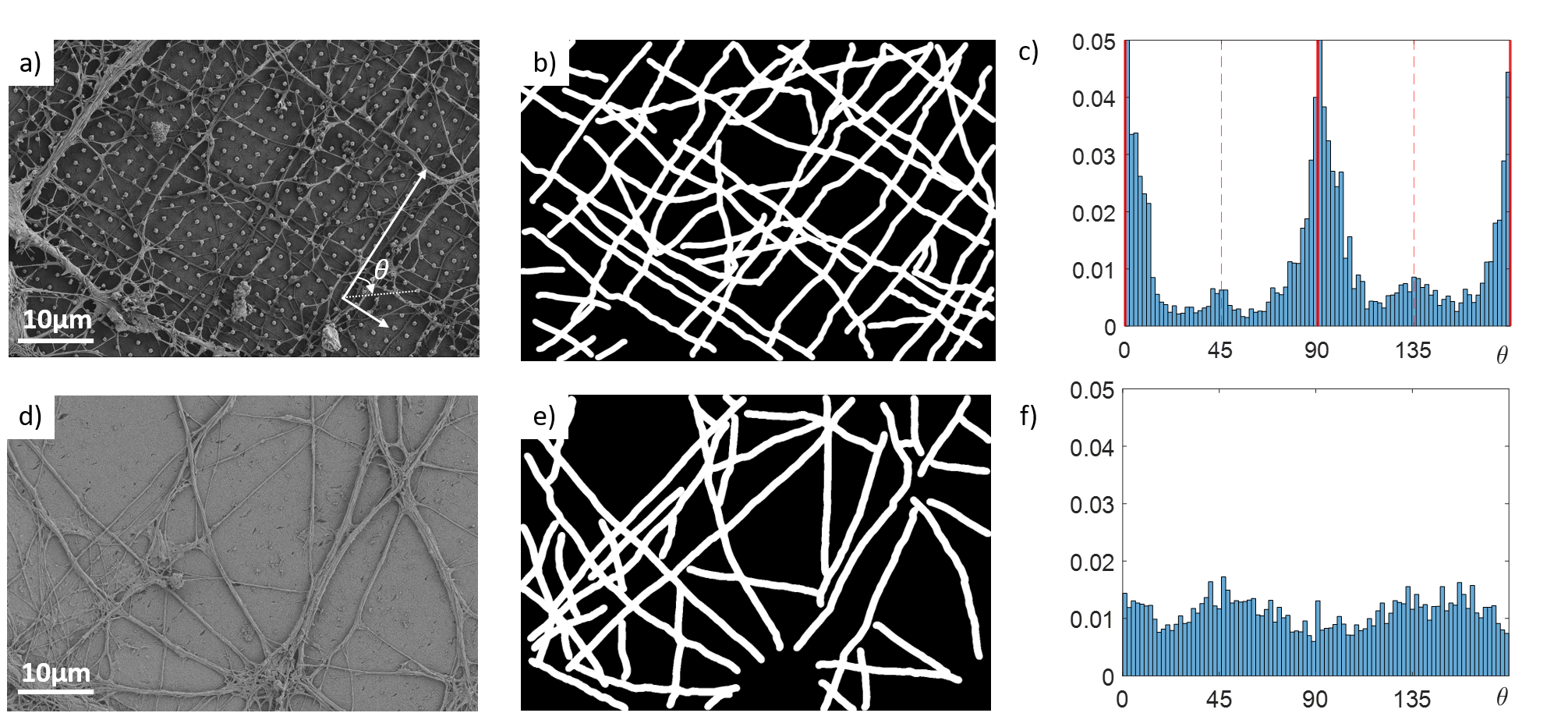}
    \caption{Preferential growth direction \red{of primary hippocampal neurons} neurites on nanopillar arrays. The first row (a,b,c) refers to a nanostructured substrate ($d=300$~nm, $p=2~\mu$m), the second row (d,e,f) to a flat surface. a) and d) SEM images at $\times$4.5k magnification. b) and e) Neurites detected using NeuronJ. c) and f) Histograms reporting the axon direction with respect to one of the main axis of the grid, as illustrated in panel a). Red lines mark the two perpendicular directions along nearest neighbors of the array. Each histogram is obtained analysing 15 images.}\label{fig:NeuronJ}
\end{figure}

We further employed digital image analysis to quantify preferential growth of the \red{neurites} along the nanopillar grid -- an effect already apparent when comparing Fig.s~\ref{fig:SEMneuronsposition}e and f. We performed computer-assisted image recognition with NeuronJ \cite{meijering2004design}, an ImageJ plugin  facilitating the tracing and measurement of elongated patterns -- here the neurites. Figs.~\ref{fig:NeuronJ}b,e show examples of output for two of the processed SEM images (original images reported in Fig.s~\ref{fig:NeuronJ}a,d). We then analyzed the output images with Python, using the Hough Line Transform to detect straight lines \cite{van2014scikit}. The slope of the detected lines is evaluated and finally, the histograms reported in Fig.s~\ref{fig:NeuronJ}c,f are obtained. \red{In order to have a statistically relevant sample size, at least 10 images (corresponding to a total estimated neurites length on the order of 1~mm) are analysed for each histogram. }

The red vertical lines in Fig.~\ref{fig:NeuronJ}c correspond to the two main directions of the pillar arrays (along the nearest neighbors), independently evaluated. This histogram presents two clear peaks along these directions, from which we can conclude that the neurites preferentially grow along the two main axes of the grid. Two smaller peaks can be identified at $\pm 45^\circ$ from the main peaks. 
These findings can be explained by the tendency of neurites to reach toward the nearest neighboring pillar when they grow from pillar to pillar. 

Finally, we confirm that on a flat, unpatterned substrate the axon orientation histogram is compatible with a uniform distribution, Fig.~\ref{fig:NeuronJ}f.
The same analysis was repeated for different geometrical parameters, \red{the corresponding histograms are reported in the Supplementary Material, Fig. \ref{fig:AllHist}. We observe that the peaks are less prominent the higher the distance between the pillars; when $p\geq 6~\mu$m the peaks vanish. In this case many neurites lie on the diamond plane below the nanostructures and their growth is virtually unaffected by the presence of nanopillars. On the opposite side, for the smallest pitch ($p=1\mu$m) the preferential growth is also attenuated with respect to the optimal case of $p=2\mu$m, which can be due to the pitch size approaching the neurites diameter.}

These findings also qualitatively agree with previous studies analyzing, with different tools, the effect of micro- and nanostructured substrates on neuronal growth, considering different cell models and different substrate materials and topographies \red{ \cite{marcus2017interactions,simitzi2017controlling,martinez2009effects,hoffman2010topography,repic2016characterization,dowell2004topographically,specht2004ordered,park2016control,piret2015support,gautam2017engineering}}.

\subsection{Neuronal activity} 

To investigate whether the nanostructured diamond surface compromises the normal neuronal activity, we carried out patch clamp recordings in a whole-cell configuration. To compare the electrophysiological properties and to determine if the geometrical parameters  (i.e. $d$ and $p$) have a significant impact on the neuronal functionality, we probed neurons (i.e. primary hippocampal neurons) from different arrays. 

\begin{figure}
    \centering
    \includegraphics[width=0.7\textwidth]{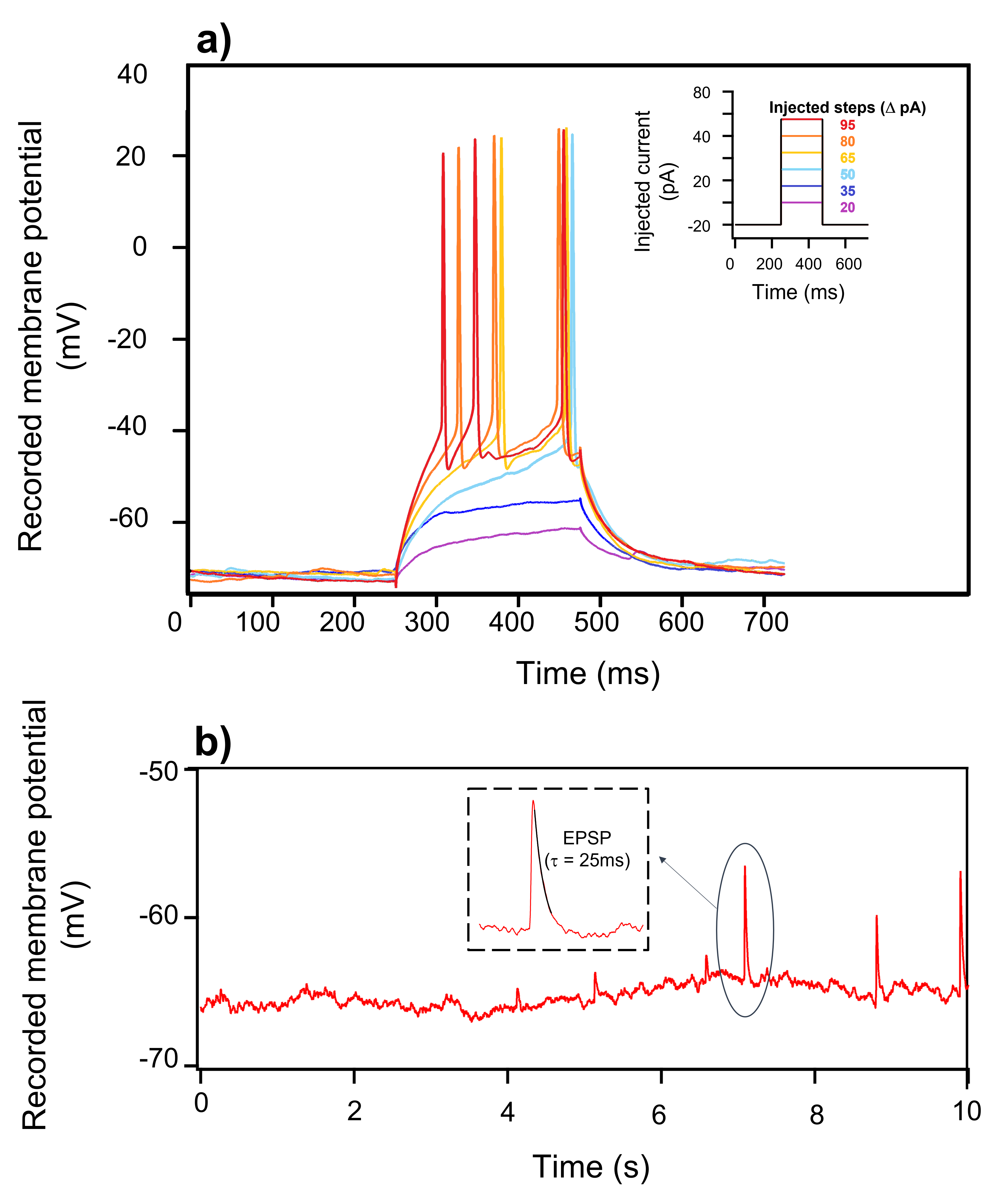}
    \caption{a) Elicited action potentials from primary hippocampal neurons plated on the nanostructured diamond substrate. The curves are obtained by injecting discrete, incremental, depolarizing current pulses close to threshold. Action potentials present a clear overshoot and hyperpolarization phase. b) Excitatory post synaptic potentials (EPSPs) produced at -65~mV membrane potential by spontaneous synaptic activation. \red{In the inset a single ESPS peak is fit giving a time decay constant $\tau \sim 25$ ms.}}\label{fig:ActionPotential}
\end{figure}

Fig.~\ref{fig:ActionPotential}a shows whole-cell recordings in the patch-clamp configuration described in the Methods section, from a neuron on the $d=400~\mathrm{nm}, p=2~\mu\mathrm{m}$ array. The intracellular current injection in subsequent steps of 15~pA amplitude and 225~ms time duration and the simultaneous recording of the membrane potential allows evaluation of the threshold current above which action potential firing is observed. This value is found to be around 30~pA. 
We recorded for several minutes without observing any decrease in the spikes amplitude.
Moreover, the spikes reported in Fig.~\ref{fig:ActionPotential}a present the typical shape of intracellular action potentials, with a clear overshoot and hyperpolarization phase \cite{bean2007action, barnett2007action}. 
The same experiment is repeated on several neurons (pillar arrays with different geometrical parameters and flat sample) always confirming the results reported in Fig.~\ref{fig:ActionPotential}a. 

The average cell membrane resting potentials in the tested arrays are \red{always between -63 mV and -49 mV. The collected results are reported in the Supplementary Material.}
We cannot establish a clear correlation between geometrical parameters (pillar diameter or pitch) and resting membrane potential. 
It is likely that differences in resting potential arise from small inhomogeneities in sample cleaning, plating, coating and patching. Remarkably, all measured values are typical of functional and healthy neurons \cite{bean2007action, barnett2007action}. 

The signal recorded in Fig.~\ref{fig:ActionPotential}b is obtained at an assigned resting membrane potential of -65mV (-40pA). The presence of spikes thus demonstrates spontaneous electrophysiological spiking activity, due to signaling from adjacent neurons. \red{In the inset an exponential fit for a single excitatory postsynaptic potential (EPSP) is reported. The resulting time decay constant ($\tau \sim 25$ ms) is within the range found in physiology \cite{fricker2000epsp}. The observed amplitude is higher than the typical values reported in literature, which may be caused by the neurons not being in physiological conditions. A more detailed discussion is provided in the Supplementary Material.} 

\section{Conclusions and outlook}

Our fabrication process allows to produce, in a reliable way, large-scale diamond nanopillar arrays in single crystal CVD diamond, which can be commercially obtained as high purity material (`electronic' or `quantum' grade).
We considered pillars of minimum diameter 100~nm and maximum height of 2~$\mu$m, and we obtained uniform arrays containing more than $\sim 10^6$ pillars over a region up to 2~mm~$\times$~2~mm, only limited by the diamond chip size. 
Using an home-built confocal microscope, we measured both the PL and ODMR spectra inside and outside the pillars, demonstrating their waveguiding effect. \red{We emphasize that being able to enhance optical excitation and collection efficiency is particularly important in the biological context, since it allows to keep the light power sufficiently low, thus avoiding phototoxicity \cite{cole2014live}.}
For the present study, we used a commercially available optical grade CVD diamond, typically containing a uniform density of NV centers (estimated as $1.4$~ppb), which is not optimal for taking advantage of the pillar geometry and for achieve high sensitivity.
NV centers should instead be present only near the top of the pillars, i.e. as close as possible to the neurites. 
To this end, one can use more costly electronic grade substrates (containing a negligible amount of NV centers) with a thin layer of NVs implanted just below the surface \cite{wrachtrup2013nitrogen, mccloskey2020enhanced}. 
Instead of ion-implantation, using a delta-doped overgrown layer may offer better NV coherence times and is within reach of present technologies \cite{ohno2012engineering}.
We could not explore the role of the functionalization layer (poly-l-lysine and laminin) on the NV coherence properties, due to the uniform distribution of NV centers in the diamond volume. This aspect will be relevant when working with shallow NV layers and should be systematically investigated in future works.
Different geometries rather than cylindrical pillars can be explored in order to increase the light collection efficiency and to promote neuronal growth and adhesion to the nanostructures. Among others, possible options are parabolic structures \cite{hedrich2020parabolic} and pyramids \cite{batzer2020single}.

Using SEM imaging and electrophysiology measurements, we showed that primary murine hippocampal neurons can be plated and are functionally active on the nanostructured diamond surfaces.
Neuron viability and function do not show a clear dependence on the geometrical parameters of the nanostructures, which is an encouraging result. 
In line with previous results \cite{gautam2017engineering, hanlon2020diamond}, we confirm that, given a suitable spacing between them, the pillars induce a preferential orientation onto the \red{neurites}' growth direction. Moreover, we conclude that, for sufficiently small distances between pillars ($p\leq 4~\mu$m), the \red{neurites} are mainly suspended and they are typically in contact with the pillar top surfaces or with the upper part of the side wall.
These result are particularly relevant \red{in the light of the recently published proof of principle experiment \cite{mccloskey2022diamond} where a similar diamond platform was used to detect voltages comparable to those generated by neuronal activity assuming that neurons would maintain their functionalities despite the nanostructuring and liying on top of the pillars. Our work demonstrates that the envisioned platform is compatible with functional neurons, which is a milestone towards an NV-based sensing platform for living neurons.}

\section{Methods}

\subsection{Diamond platform nanofabrication}\label{Sec:fabrication}

\begin{figure}
    \centering
    \includegraphics[width=0.9\textwidth]{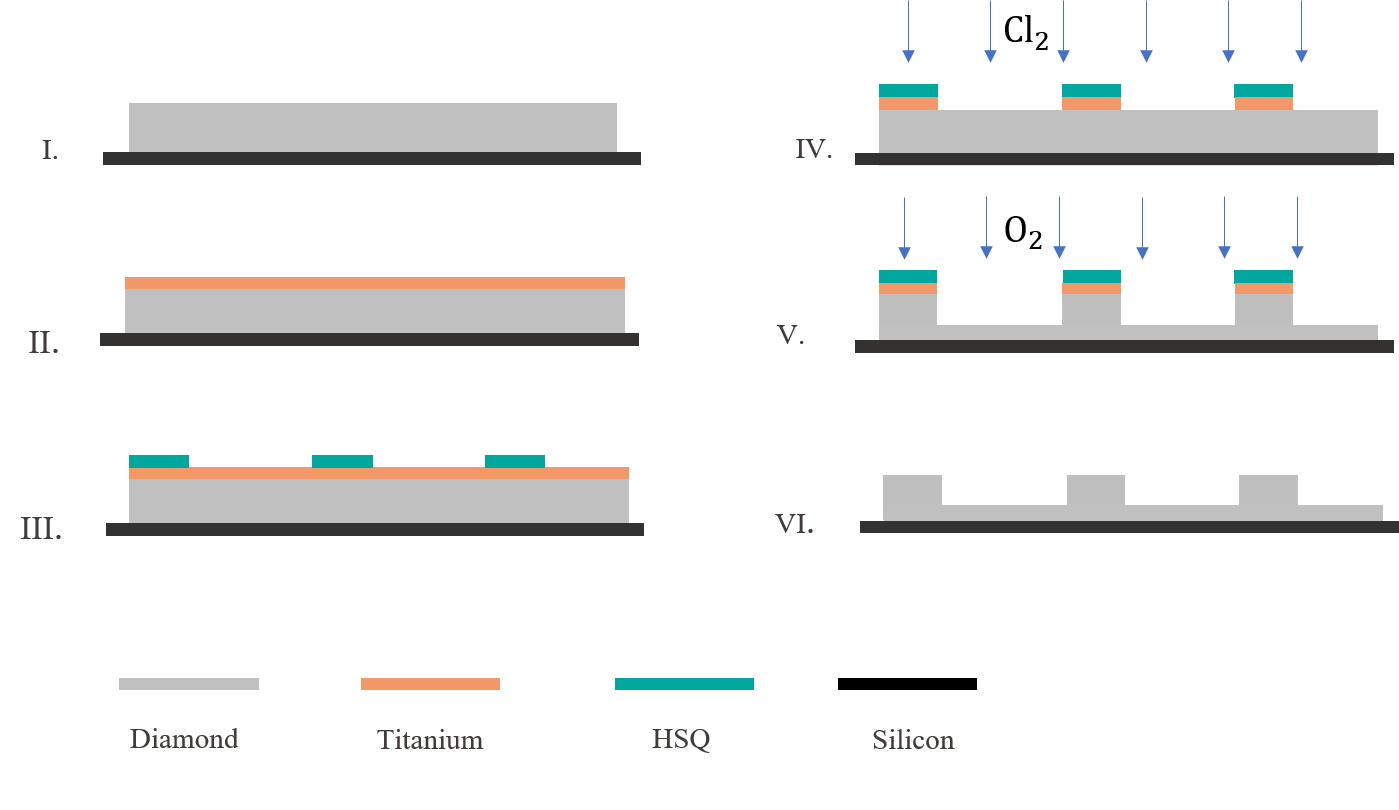}
    \caption{Fabrication process for large-scale diamond nanopillar arrays. I. The single-crystal diamond substrate is cleaned, polished (see main text) and die-attached on a Si carrier  wafer. II. Sputtering of Ti hard mask (200 nm). III. Spin-coating of HSQ negative resist (150 nm), electron beam lithography for pillars, and development in TMAH. IV. $\mathrm{Cl}_2$-based reactive ion etching patterning of the Ti layer. V. Highly directional $\mathrm{O}_2$-plasma etch of diamond substrate. VI. Stripping of the hard mask in diluted hydrofluoric acid (HF).}\label{fig:ProcessFlow}
\end{figure}

We fabricate our nanopillar arrays on commercial $\langle 100 \rangle$ CVD single crystal diamond substrates (3mm x 3mm x 0.25mm, optical grade, Element6). The nanofabrication process flow is schematically reported in Fig.~\ref{fig:ProcessFlow}. The sample is mounted via an adhesive (QuickStick mounting wax) on a standard Si wafer, for easier handling and processing. It is then cleaned first in acetone, and then in piranha solution (3:1, $\mathrm{H}_2 \mathrm{SO}_4:\mathrm{H}_2\mathrm{O}_2$, 3 min) to remove any contamination. In order to remove residual mechanical polishing lines and other surface defects, a  preliminary non-contact polishing step is applied. This process is based on physical bombardment of the diamond surface with accelerated inert gas ions and has been described in detail in \cite{mi2019non}. After our non-contact polishing, a root mean square (rms) roughness of $<2$~nm is measured over a $100~\mu\mathrm{m}^2$ region. Note that a sufficiently low surface roughness is crucial for obtaining nanostructures of high quality.
After evaporation of 200~nm of Ti and spin coating of $\sim$150~$\mu$m Hydrogen silsesquioxane (HSQ XR-1541-006) negative resist, the pillars are patterned using electron beam lithography and subsequent development in Tetramethylammonium hydroxide (TMAH 25\%). The typical exposure time is less than 20 minutes for a  2~mm~$\times$~2~mm array. This approach provides both precise and tunable size and shape control. 

In order to transfer the pattern from the resist to the titanium hard mask and to diamond several dry etching process steps are applied. First, a $\mathrm{Cl}_2$-based reactive ion etching process allows to pattern the Ti layer (STS Multiplex ICP, 800~W ICP power, 150~W bias power, 10~sccm $\mathrm{Cl}_2$ and 10~sccm $\mathrm{BCl}_3$, 3~mTorr). Second, to vertically etch the diamond a highly directional $\mathrm{O}_2$-plasma etch is used, resulting in an etch rate of 100~nm/min  (STS Multiplex ICP, 400 W ICP power, 200~W bias power, 30~sccm $\mathrm{O}_2$, 15~mTorr). A comprehensive review of different recipes for reactive ion etching of single crystal diamond by inductively coupled plasma can be found in \cite{toros2020reactive}. Finally, the HSQ and Ti masks are removed by diluted HF etch (1\% concentration)

Using this process, we successfully fabricate square arrays of pillars of diameter ($d$) between 100~nm and 500~nm, with a height ($h$) up to 2~$\mu$m, homogeneously covering areas as wide as 2~mm~$\times$~2~mm, only limited by the diamond sample size. Aspect ratios ($h:d$) exceeding 10:1 are obtained. The distance between the pillars (pitch, $p$) can be adjusted by design and in this experiment is varied between 1 and 10~$\mu$m. 

The height and diameter of the pillars are chosen in accordance to simulation studies which conclude that those values are particularly suitable for enhancing photoluminescence waveguiding effect \cite{hanlon2020diamond,fuchs2018optimized}. The distance between the pillars is varied in a range that still allows sub-cellular spatial resolution (typical mammalian neurons have a cell body diameter of few tens of $\mu$m). 

 Due to the sample preparation protocol outlined above, the optimized microfabrication process results in a particularly smooth etched diamond bottom surface, and the pillars present a smooth and flat top with highly vertical sidewalls, as revealed in the SEM recordings of individual pillars and large-scale arrays, Fig.~\ref{fig:Substrate}. Our fabricated arrays consist routinely in more than $10^6$ individual nanopillars, and we observe no significant failure.
 
\subsection{PL and ODMR measurements}
Both photoluminescence and ODMR measurements were carried out in our home-built confocal fluorescence microscope. More information about the setup and the acquisition software can be found in \cite{babashah2022optically}. In order to achieve sufficiently high spatial resolution we use a high numerical aperture objective (N.A.= 0.9) and we collect the signal with a single mode fiber and a single photon detector. The excitation laser power was $\sim 5$mW before the objective, the wavelength $\lambda=515$nm.
For PL maps, we scan the sample using a nanopositioning stage. The acquired $xy-$ and $xz-$ maps have a resolution of 20nm. In Fig.s~\ref{fig:PL}a-b a median filter ($3 \times 3$ window) is applied to the raw data. Considering the diamond index of refraction $n=2.4$, the total scanning depth inside the diamond is approximately $10 \mu$m, i.e. much greater than the pillar height $h\sim 1\mu$m. 
For ODMR, microwaves (26 dBm) are applied using an antenna patterned on a PCB board, offering great uniformity across the sample \cite{sasaki2016broadband}. In order to observe the four Zeeman splittings, we apply an external magnetic field using a permanent magnet ($\sim 3-4$ mT). The direction of the field was adjusted to split the degeneracy and distinguish all the 8 deeps, corresponding to the field component along the 4 possible NV-axis directions. For convenience, just half of the spectrum is reported in Fig.s~\ref{fig:PL}d-e, the other 4 deeps are symmetrical respect the zero-field value 2.87 GHz.
For both PL and ODMR spectra, typical acquisition times are of the order of few minutes.

\subsection{Cell culture plating and maintenance}

Primary hippocampal neurons were prepared from P0/P1 pups from WT C57BL/6JRj mice, cultured and maintained according to a previously published protocol \cite{steiner2002overexpression}. To facilitate the growth of neurons on the diamond surface, we coated the diamond chips with poly-l-lysine (Bio-Techne, 3438-100) and laminin (SIGMA-ALDRICH, L2020). More specifically, sterile diamond chips were placed in a well of a 96 well plate, coated with poly-l-lysine for one-hour at room temperature and then washed 3 times with distilled water. Subsequently, the diamond chips were coated with 50~$\mu$l/ml laminin and incubated for about 30 minutes at 37$^\circ$C. Finally the plates were rewashed in distilled water for 3 times and kept for drying. We found that using only poly-l-lysine or only laminin prevents proper growth of neurons.

The primary hippocampal neurons were plated at a density of 150'000~cells/ml and in each well (one for each diamond chip considered) about 200~$\mu$l of medium was added (Neurobasal medium [Thermofisher, 21103049], 2\% B27 supplement [ThermoFisher, 17504044], 50~mM L-Glutamine, 1\% Penicillin/Streptomycin). The neurons were maintained in an incubator at 37$^\circ$C with 5\% $\mathrm{CO}_2$ for 10-14 days before measurements or imaging.
In order to minimize the differences among the samples, the diamond chips used for an experiment were always processed at the same time, and in the same 96 well plate. In addition to the nanostructured diamond chips, also a flat diamond is considered as control.

\subsection{Scanning Electron Microscopy (SEM)}

Scanning electron microscopy (SEM) was employed to examine the diamond nanostructured surface topography and to evaluate cell morphology and growth on the different diamond chips (both nanostructured and flat). 
No specific chip preparation was necessary to acquire SEM images of the pillar array without cells (Fig.~\ref{fig:Substrate}). 

Primary neurons (10 DIV) plated on the diamond substrates (both flat and nanostructured) were prepared for SEM as follows: (0) PBS rinsing to get rid of any proteins in the solution ($2\times 3$~s); (1) fixation with 1.25\% glutaraldehyde in phosphate buffer (0.1M, pH 7.4) (1-2~hours); (2) washing in cacodylate buffer (0.1M, pH7.4) ($3\times 2$~ min); (3) postfixation in 0.2\% $\mathrm{OsO}_4$ (30~min); (4) rinsing in deionized water ($2\times 3$~min); (5) immersion in increasing concentrations of alcohol: $1\times 30$\% $1\times 50$\%, $1\times 70$\%, $1\times 90$\%, $1\times 96$\%, $2\times 100$\% (3~min each) (6) critical point drying (7) evaporation of 4~nm of Au/Pd layer on the sample surface. A thickness of only 4~nm is allowed by the $\mathrm{OsO}_4$ treatment, moreover, the gold-palladium offers a less granular coating with respect to only gold \cite{brunk1981fixation}. 

All SEM images were acquired by using an ultra-high resolution FE-SEM (field emission scanning electron microscope, Zeiss Merlin, GeminiII column). Typical working parameters are: operating voltage 1.50~kV, current 80~pA, 
stage tilting angle between 0$^\circ$ and 30$^\circ$.
For imaging the diamond chips before plating the cells lower current (20~pA) was used in order to reduce the charge-induced drifting effect.

\subsection{Electrophysiology}

Primary neurons cultured on a nanostructured diamond chip were placed after 14 days in vitro (14 DIV) in a recording chamber at room temperature while gently perfused by oxygenated (95\% $\mathrm{O}_2$, 5\% $\mathrm{CO}_2$) extracellular solution (in mM):  125 NaCl, 2.5 KCl, 1 MgCl $\cdot$ 6$\mathrm{H}_2\mathrm{O}$, 1.25 $\mathrm{NaH}_2\mathrm{PO}_4$, 2 $\mathrm{CaCl}_2$ $\cdot$ H2O, 25 Glucose, 25 $\mathrm{NaHCO}_3$. 
Whole cell recordings in current clamp mode were performed under the guidance of a motorized (SM-5 controller, Luigs – Neumann) IR-DIC microscope (BX51WI, Olympus) with a water immersion 40X objective, and a PRIME 95b camera (Teledyne photometrics). 
The recording Ag/AgCl electrodes were fabricated with borosilicate glass pipettes (P97, Sutter Instrument) of 7 to 10 $\mathrm{M}\Omega$ resistance and filled with the following intracellular solution (in mM): 110 K-gluconate, 10 HEPES, 10 KCl, 4 Mg-ATP, 0.3 Na-GTP, 10 Na-phosphocreatine, Mannitol to adjust osmolarity to 290 mOsm, and KOH to adjust pH 7.2.
The signal was amplified with an Axopatch 200B amplifier (Axon Instruments), filtered (low pass 1kHz), digitized (20 kHz), and acquired through an ADC/DAC data acquisition interface (ITC-1600, Instrutech) by using a custom-made software running on Igor Pro (Wavemetrics). 
Resting membrane potential was obtained at 0pA. Current was injected to preserve membrane potential at -65mV and recorded to observe spontaneous excitatory postsynaptic potentials (EPSPs). Action potentials were elicited by successive, discrete incremental current pulses of 225 ms.

\section{Acknowledgements}
This project has received funding from the EPFL Interdisciplinary Seed Funds and from the Swiss National Science Foundation under grant \#183717 and \#198898.
We sincerely thank the EPFL Center for Imaging (especially Wittwer Mallory) for their useful insights on image processing with ImageJ, Panagaki Theodora for her help in culturing the cells, Knott Graham and the BioEM staff for their expertise in cells preparation for SEM imaging. The diamond nanopillar arrays were fabricated at the EPFL Center of MicroNanoTechnology (CMi) and we gratefully acknowledge the support of EPFL CMi.

\bibliography{Ref}

\newpage
\section{Supplementary Material}

\renewcommand{\thefigure}{S1}
\subsection{Histograms for the different pitches}
\begin{figure}
    \centering
    \includegraphics[width=1\textwidth]{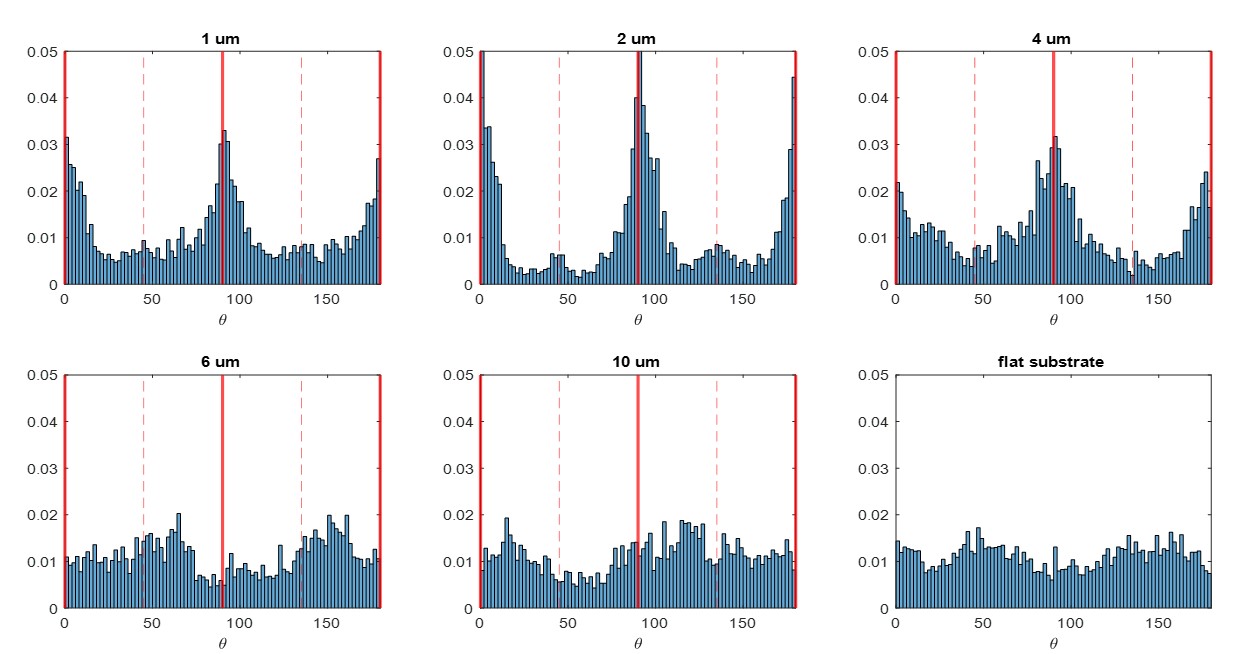}
    \caption{Histograms reporting the axon direction with respect to one of the main axis of the grid, for all the pitches present in our sample design. The detail of the analysis can be found in the main text. Red lines mark the two perpendicular directions along nearest neighbors of the array. Red dashed lines correspond to 45 and 135. Each histogram is obtained analysing a number of SEM images between 10 and 15.}\label{fig:AllHist}
\end{figure}

\subsection{Preliminary results on neuronal culture on diamond substrate}

We report here the results from the first test we did about culturing neurons on diamond. As reported in literature, a thin functionalization layer is typically necessary to promote adhesion between the cells and the diamonds. In particular we tried:
\begin{itemize}
    \item poly-l-lysine (Bio-Techne, 3438-100) 
    \item laminin (SIGMA-ALDRICH, L2020)
    \item poly-l-lysine (Bio-Techne, 3438-100) and laminin (SIGMA-ALDRICH, L2020)
\end{itemize}
Only the combination of the two was successful, while using just one of the coating resulted in poor quality of the culture (visible already from optical microscope inspection).

For the final experiment we plated at a density of 150’000 cells/ml and we imaged the neurons after 10 days, with the results reported in the main text. Preliminary attempts considered higher density (2x) and longer incubation time, up to 15 days. In this cases, we observed that it was not possible to see the diamond nanostructures, since they were completely embedded in the culture. An example of this situation is reported in the following figure.

\renewcommand{\thefigure}{S2}
\begin{figure}
    \centering
    \includegraphics[width=1\textwidth]{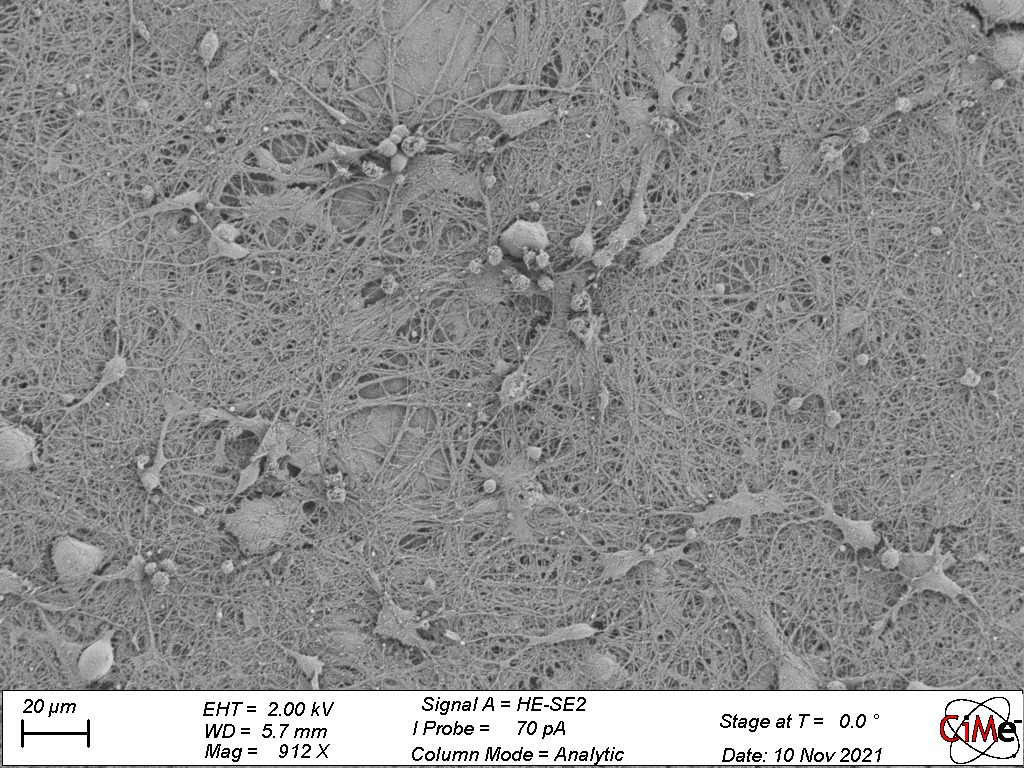}
    \caption{SEM image of primary hippocampal neuron plated on nanostructured diamond, after 14 days of incubation and with higher cell concentration respect to what was chosen for the final experiment. }\label{fig:HighDensitySEM}
\end{figure}

\subsection{Resting membrane potential}
In the following table we report the average cell membrane resting potentials measured in the different tested arrays.
\begin{center}
\begin{tabular}{ c c }
 Resting membrane potential (mV) & $p$($\mu$m) $\times$ $d$(nm) \\ 
 -55 & 4 $\times$ 300 \\  
 -60 & 1 $\times$ 500 \\  
 -61 & 1 $\times$ 400 \\  
 -49 & 10 $\times$ 100 \\  
 -49 & 6 $\times$ 200 \\  
 -63 & 2 $\times$ 200 \\ 
 -60 & flat
\end{tabular}
\end{center}

\subsection{On the amplitude of EPSPs recordings}

As mentioned in the text the amplitude of the EPSPs signal we recorded (10 mV) is higher compared to typical values reported in literature. We here discuss better this point, suggesting some hypothesis explaining the phenomenon.

First of all, we need to take into account that the neurons are not in physiological conditions, and their neurites are not receiving the same signals for connecting than in physiology. It is possible that anomalous (in this case more than usual) connectivity happens between 2 neurons. 
It is possible that, due to the conditions of the culture (smaller neurite growth and higher interconnectivity), the amplitude of EPSP is observed larger than the value typically recorder for neurons in the physiology of brain tissue. Also, for the cell corresponding to the measurement reported in the figure, we were injecting -40pA to keep the recording at -65mV. Injection of negative current might have and amplifying effect on the amplitude of EPSPs.

\end{document}